\documentstyle[aps,epsf,tabularx,multicol]{revtex}
\begin{document}
\tighten
\title{Distribution functions and current-correlations\\ in
normal-metal---superconductor hetero-structures}
\author{Thomas Gramespacher and Markus B\"uttiker}
\address{D\'epartement de Physique Th\'eorique, Universit\'e de Gen\`eve,
CH-1211 Gen\`eve 4, Switzerland}
\date{\today}
\maketitle
\begin{abstract}
We introduce electron-like and hole-like distribution functions,
which determine the currents and the fluctuation spectra of the
currents measured at a normal-conductor---superconductor
hetero-structure. These distribution functions are expressed with
the help of newly defined partial densities of states for
hetero-structures. Voltage measurements using a weakly coupled
contact on such a structure show the absence of a contact
resistance to the superconducting reservoir and illustrate how the
interface to the superconductor acts as an Andreev mirror. We also
discuss the current-current correlations measured at two normal
contacts and argue that the appearance of positive correlations is
a purely mesoscopic effect, which vanishes in the limit of a large
number of channels and in the average over an ensemble.\\[0.3cm]
\end{abstract}
\begin{multicols}{2}
\narrowtext
\section{introduction}
The properties of a phase-coherent normal conductor can be
drastically changed by the presence of a nearby superconductor
($N-S$ system). At the interface to the superconductor electrons
with energies smaller than the superconducting gap are Andreev
reflected\cite{andreev,blonder.5} and scattered back as holes,
thus inducing superconducting behavior in the normal conductor.
This process is known under the name {\em proximity effect} and
has during the last few years extensively been studied
experimentally \cite{petrashov,takayanagi,courtois,gueron,hartog}
and theoretically \cite{beenakkerreview}. In particular, Gu\'eron
{\em et al}.\cite{gueron} succeded in measuring the opening of a
gap in the local density of states of a normal conductor in the
neighborhood of a superconductor using tunneling into a spacially
extended contact. In the theoretical treatment, most often, the
influence of the superconductor on a disordered normal conductor
is studied using a semi-classical Green's function
technique\cite{wilhelm}. We investigate here phase-coherent $N-S$
structures in a fully quantum-mechanical framework and show the
influence of the superconductor on measurements done at a normal
contact, which is weakly coupled to the hetero-structure. We give
a quantum-mechanical expression of the particle and hole
distribution functions of a structure that can be either
ballistic, containing few scatterers (clean) or disordered and
treated in the ensemble average. These expressions are based on
newly defined partial densities of states for hybrid
structures\cite{gramespacherthesis}. One of the most striking
effects is that the correlations of currents at two normal
contacts can become positive due to the bosonic correlations
induced in the normal conductor by the nearby
superconductor\cite{anantram.5,martin.5,torres.5}. We present a
geometry that leads to positive correlations and argue that such
positive correlations are truly mesoscopic and will vanish if one
goes to the limit of a large number of channels or performs
averages over disorder.
\begin{figure}[htbp]
\epsfxsize7.5cm \centerline{\epsffile{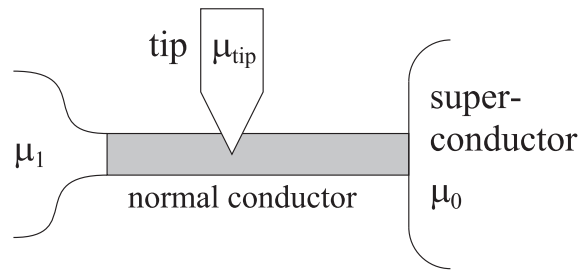}}
\vspace{0.15cm} \caption{A normal-conductor---superconductor
hetero-structure. The normal conducting side has two terminals,
one of which, the tip, is only weakly coupled to the rest of the
system. The gray shaded area can be disordered or ballistic.}
\label{nsfig.5}
\end{figure}
\section{The effective scattering matrix}
We consider an $N-S$ hetero-structure consisting of a normal
conducting part connected to a superconductor. The superconductor
represents one terminal whereas the normal conducting part is
connected to $N$ normal conducting terminals. In Fig.\
\ref{nsfig.5} such a structure with $N=2$ normal contacts is drawn
schematically. One of the normal contacts is only weakly coupled
to the rest of the system just next to the interface to the
superconductor. Such a setup permits to compare local properties
of a normal wire connected to two normal reservoirs with the
properties of a normal wire connected to one normal and one
superconducting reservoir. An additional normal contact on a
heterostructure has also been used by Mortensen {\em et
al}.\cite{mortensen} to investigate dephasing, thus generalizing
B\"uttiker's model\cite{buttiker85} based on dephasing reservoirs
for heterostructures.

At applied potentials and temperatures much smaller than the gap
of the superconductor, we are interested in the scattering
properties of electrons injected through one of the normal
contacts of an $N-S$ hetero-structure as e.\ g.\ the one shown in
Fig.\ \ref{nsfig.5}. The normal part can be disordered and is
described by the self-consistently calculated equilibrium
electrostatic potential $U(x)$. The Hamiltonian describing the
normal part of the hetero-structure can be written as a $2\times
2$ blockmatrix $\widehat{H}$, which acts in the combined particle
and hole space (Nambu space),
\begin{equation}
\widehat{H}=\left(
\begin{array}{cc}
-\frac{\hbar^2}{2m}\nabla^2+q^eU^e(x) & 0 \\ 0 &
\frac{\hbar^2}{2m}\nabla^2+q^hU^h(x)
\end{array}
\right)\, .
\end{equation}
Here, we introduced the electron charge $q^h=e$, the hole charge
$q^h=-e$ and the electron and hole potentials
$U^e(x)=U^h(x)=U(x)$.

Let us first consider the case, where the superconductor is
replaced by a normal conductor. The Green's function of the
disordered part containing the coupling to the $N+1$ contacts and
the scattering matrix describing the scattering of particles
between the $N+1$ contacts can then be written in Nambu space as
$2\times2$ block-diagonal matrices\cite{weidenmueller},
\begin{eqnarray}
\widehat{G} & = &
(E-\widehat{H}+i\pi\widehat{\Gamma})^{-1}\label{gnormal.5}\, , \\
\widehat{U} & = & 1-2\pi i \widehat{W}^\dagger
\widehat{G}\widehat{W}\, . \label{unormal.5}
\end{eqnarray}
We extended the coupling matrix $W$ for electrons to a coupling
matrix for electrons and holes
\begin{equation}
\widehat{W}=\left(
\begin{array}{cc}
W & 0 \\
0 & W^*
\end{array}
\right)\, ,
\end{equation}
and $\widehat{\Gamma}=\widehat{W}\widehat{W}^\dagger$. The Green's
function $\widehat{G}$ and the scattering matrix $\widehat{U}$ are
block-diagonal matrices. In a purely normal-conducting system
there is no mixing between particle and hole space.

The presence of a superconductor mixes now these two spaces.
Incoming particles can be reflected as holes and vice versa. We
assume that the interface between the normal conductor and the
superconductor is perfectly transparent. All disorder is contained
in the normal part of the system. The scattering between particles
and holes at the interface is at the Fermi energy (the chemical
potential of the condensate in the superconductor) described by
the Andreev-reflection matrix \cite{blonder.5}
\begin{equation}
R_A=\left(
\begin{array}{cc}
0 & -i \\
-i & 0
\end{array}
\right)\, .
\end{equation}
This is a $2\times 2$ block-matrix with each block being a diagonal matrix of
the size of the number of channels at the interface.
Putting the scattering matrix $\widehat{U}$ of the normal region together with the
Andreev reflection matrix gives the effective scattering matrix
\begin{equation}
\widehat{S}=\left(
\begin{array}{cc}
S^{ee} & S^{eh} \\
S^{he} & S^{hh}
\end{array}
\right)
\end{equation}
describing the scattering between the $N$ normal contacts. We now
divide the scattering matrix
\begin{equation}
\widehat{U}=\left(
\begin{array}{cc}
\widehat{U}_{00} & \widehat{U}_{01} \\
\widehat{U}_{10} & \widehat{U}_{11}
\end{array}
\right)
\end{equation}
in such a way that the index 0 denotes the electron and hole
channels at the superconducting interface and the index 1 denotes
all channels of all $N$ normal contacts. The effective scattering
matrix can then be written in the following compact form,
\begin{equation}
\widehat{S}=\widehat{U}_{11}+\widehat{U}_{10}(1-R_A\widehat{U}_{00})^{-1}R_A\widehat{U}_{01}\, .
\end{equation}
This effective scattering matrix is now not anymore diagonal, but
contains sub-matrices $S^{he}$ and $S^{eh}$ which describe
scattering of incoming particles which are reflected as holes and
incoming holes reflected as particles. It can be shown that we can
also express this effective scattering matrix in the same form as
in Eqs.\ (\ref{gnormal.5}) and (\ref{unormal.5}),
\begin{eqnarray}
\widehat{S} & = & 1-2\pi i \widehat{W}_1^\dagger \widehat{G}_{eff}
\widehat{W}_1\, ,\\ \widehat{G}_{eff} & = &
(E-\widehat{H}_{eff}+i\pi\widehat{\Gamma}_1)^{-1}
\end{eqnarray}
by introducing the effective Hamiltonian\cite{frahm.5}
\begin{equation}
\widehat{H}_{eff}= \left(
\begin{array}{cc}
H & -\pi W_0W_0^t \\
-\pi (W_0W_0^t)^* & -H^*
\end{array}
\right)
\end{equation}
and where we divided the coupling matrix $W$ up into a part $W_0$
describing the coupling to the superconducting contact  and a part
$W_1$ describing the coupling to all normal contacts. We now
define the fundamental local partial densities of states for an
$N-S$ hybrid structure in analogy to the definition for normal
systems in \cite{gasparian} as
\begin{eqnarray}
\nu(\alpha_\mu,x_\nu,\beta_\lambda) = & & -\frac{1}{4\pi i}{\rm
Tr}\left\{\left( s_{\alpha\beta}^{\mu\lambda}\right)^\dagger
\frac{\delta s_{\alpha\beta}^{\mu\lambda}}{q^\nu\delta
U^\nu(x)}\right.\nonumber \\ & & \left.-\frac{\delta\left(
s_{\alpha\beta}^{\mu\lambda}\right)^\dagger}{q^\nu\delta
U^\nu(x)}s_{\alpha\beta}^{\mu\lambda} \right\} \, .
\label{fondpdos}
\end{eqnarray}
Here, the indices $\alpha$ and $\beta$ denote the normal contacts
of the system and the indices $\mu$, $\nu$ and $\lambda$
distinguish between electrons and holes. As an example,
$\nu(2_h,x_e,1_e)$ gives the electronic density at point $x$ of an
electron that entered the system through contact 1 and left the
system as a hole in contact 2. The usefulness of such a definition
will become clear soon and is also shown in connection with charge
fluctuations of a heterostructure in \cite{martin}. We proceed by
defining the electron like injectivity of a contact $\beta$ as
\begin{equation}
\nu(x_e,\beta_e)=\sum_{\alpha,\mu}\nu(\alpha_\mu,x_e,\beta_e)
\end{equation}
and the hole-like injectivity as
\begin{equation}
\nu(x_h,\beta_e)=\sum_{\alpha,\mu}\nu(\alpha_\mu,x_h,\beta_e)\, .
\end{equation}
The electron-like injectivity gives the electronic density inside
the conductor of an incoming electron from the reservoir and the
hole-like injectivity gives the hole density of an incoming
electron. Expressed with the help of the effective Green's
function $\widehat{G}_{eff}$ these densities are
\begin{eqnarray}
\nu(x_e,\beta_e) & = & \langle x|G_{eff}^{ee}\Gamma_\beta^e
(G_{eff}^{ee})^\dagger | x\rangle\, , \\ \nu(x_h,\beta_e) & = &
\langle x|G_{eff}^{he}\Gamma_\beta^e (G_{eff}^{he})^\dagger |
x\rangle\, .
\end{eqnarray}
In the same way, we can define electron- and hole-like
emissivities of a contact $\alpha$, $\nu(\alpha_\mu,x_\nu)$, by
summing the local partial density
$\nu(\alpha_\mu,x_\nu,\beta_\lambda)$ over the incoming contacts
$\beta_\lambda$. In the presence of a magnetic field $B$ we have
the following symmetry relations,
\begin{eqnarray}
\nu_B(x_\mu,1_\nu) & = & \nu_{-B}(1_\nu,x_\mu)\,
,\label{symrel1.5}
\\ \nu_B(x_e,1_e) & = & \nu_{-B}(x_h,1_h)\, , \\
\nu_B(x_h,1_e) & = & \nu_{-B}(x_e,1_h) \label{symrel2.5}\, .
\end{eqnarray}
The local density of states at a point $x$ is the sum
\begin{eqnarray}
\nu(x) & = &
\frac{1}{2}\sum_{\beta,\mu,\lambda}\nu(x_\mu,\beta_\lambda)
\nonumber \\ & = & -\frac{1}{2\pi}{\rm
Im}[G_{eff}^{ee}(x,x)+G_{eff}^{hh}(x,x)]\, .
\end{eqnarray}
The electronic and hole-like injectivity can also be expressed
with the help of scattering states. In the presence of
superconductivity, the wavefunction $\psi$ of an incoming electron
consists of two components, one being the electron wavefunction
$\psi^e$ and one being the hole wavefunction $\psi^h$. The
amplitude of the scattering state $\psi_{\beta n}$ describing an
incoming electron in channel $n$ of contact $\beta$ is then at a
point $x$
\begin{equation}
\psi_{\beta n}(x)=\left( \psi_{\beta n}^e(x) \atop \psi_{\beta
n}^h(x) \right) \, .
\end{equation}
The injectivities are proportional to the absolute squares of the
electron and hole wavefunction,
\begin{eqnarray}
\nu(x_e,\beta_e) & = & \sum_{n\in \beta}\frac{1}{hv_{\beta
n}}|\psi_{\beta n}^e(x)|^2 \, , \\
 \nu(x_h,\beta_e) & = & \sum_{n\in
\beta}\frac{1}{hv_{\beta n}}|\psi_{\beta n}^h(x)|^2\, .
\end{eqnarray}
With the help of the symmetry relations
(\ref{symrel1.5})-(\ref{symrel2.5}) the emisiivites can be
expressed using the scattering states of the Hamiltonian with the
reversed magnetic field.
\section{The distribution functions}
We consider now an $N-S$ structure to which on the normal
conductor a local tunneling contact is added as shown in figure
\ref{nsfig.5}. We have therefore a setup with three contacts, two
of which are normal and one superconducting contact. At small
temperatures and applied potentials $\mu_\alpha$ measured relative
to the electro-chemical potential of the superconductor the
average current at a contact $\alpha$ is\cite{lambert}
\begin{eqnarray}
\langle I_\alpha\rangle & = & \sum_\mu \langle I_\alpha^\mu\rangle
\nonumber \\ & = & \sum_\mu\frac{q^\mu}{h}\int_0^\infty dE
\sum_{\beta\nu}
T_{\alpha\beta}^{\mu\nu}[f_\alpha^\mu(E)-f_\beta^\nu(E)]\, .
\label{finitecurrfor.5}
\end{eqnarray}
The indices $\mu$ and $\nu$ distinguish between electrons and
holes and $\alpha$ and $\beta$ label the normal contacts. The
normal and Andreev transmission and reflection probabilities
$T_{\alpha\beta}^{\mu\nu}$ are calculated from the effetive
scattering matrix $\widehat{S}$,
\begin{equation}
T_{\alpha\beta}^{\mu\nu} ={\rm Tr}\left[ s_{\alpha\beta}^{\mu\nu}
(s_{\alpha\beta}^{\mu\nu})^\dagger \right]\, .
\end{equation}
The distribution functions of electrons and holes $f_\alpha^{e/h}$
injected from a normal reservoir $\alpha$ are
\begin{equation}
f^e_\alpha(E)=f_0(\mu_\alpha-E)\quad\mbox{\rm and}\quad
f_\alpha^h(E)=f_0(-\mu_\alpha-E) \label{distfunct}
\end{equation}
with the Fermi distribution function
$f_0(\varepsilon)=(\exp(-\varepsilon/kT)+1)^{-1}$. Note, that in
Eq.\ (\ref{finitecurrfor.5}) we integrate only from zero, the
electro-chemical potential of the superconductor, to infinity in
order to avoid double counting. In the absence of a superconductor
(i.\ e.\ replacing the Andreev reflection by normal reflection)
the current formula (\ref{finitecurrfor.5}) reduces to the well
known equation for purely normal systems \cite{buttiker92.5}.

For temperatures and potential differences that are much smaller
than the superconducting gap, we can neglect the energy dependence
of the Andreev reflection matrix. For the current at the tunneling
contact we need the transmission probabilities from and to the
massive contact to first order in the coupling constant $t$,
\begin{eqnarray}
T_{tip,1}^{\mu\nu} & = & 4\pi^2\nu_{tip}|t|^2\nu(x_\mu,1_\nu)\, ,
\\ T_{1,tip}^{\mu\nu} & = & 4\pi^2\nu(1_\mu,x_\nu)|t|^2\nu_{tip}\, .
\end{eqnarray}
Here, $\nu_{tip}$ is the density of states at the tip. Using these
transmission probabilities we can rewrite Eq.\
(\ref{finitecurrfor.5}) in the form of a two-terminal current,
\begin{eqnarray}
\langle I_{tip}\rangle=\frac{e}{h}\int_0^\infty dE G(x) & \left[
\right. & (f_{tip}^e(E)-f_{eff}^e(E) ) \nonumber \\ & - & \left.
(f_{tip}^h(E)-f_{eff}^h(E) ) \right]\, . \label{effcurrglg}
\end{eqnarray}
Here, we defined the effective distribution of electrons
\begin{equation}
f_{eff}^e(E) =
\sum_{\beta\nu}\frac{\nu(x_e,\beta_\nu)}{\nu(x)}f_\beta^\nu(E)
\label{effdisfunce}
\end{equation}
and of holes
\begin{equation}
f_{eff}^h(E) =
\sum_{\beta\nu}\frac{\nu(x_h,\beta_\nu)}{\nu(x)}f_\beta^\nu(E)
\label{effdisfunch}
\end{equation}
at a point $x$ inside the conductor. Due to the possibility of
Andreev reflection, injected electrons can contribute to the
distribution of holes and injected holes can contribute to the
distribution of electrons. Equations
(\ref{effcurrglg})-(\ref{effdisfunch}) are central results of this
paper and represent a generalization of similar formulas developed
in \cite{gramespacher99} for purely normal conducting systems.

At zero temperature the distribution function of electrons and
holes can be replaced by step functions,
\begin{equation}
f_\alpha^e(E)=\theta(\mu_\alpha-E)\quad\mbox{\rm and}\quad
f_\alpha^h(E)=\theta(-\mu_\alpha-E)\, .
\end{equation}
Using these distribution functions in (\ref{finitecurrfor.5}) and
neglecting the energy dependence of the transmission probabilities
the currents at the contacts are in linear response to the applied
bias determined by the transmission and reflection probabilities
at the Fermi energy,
\begin{eqnarray}
\langle I_{tip}\rangle & = & \frac{e^2}{h}(
N_{tip}-T_{tip,tip}^{ee}+T_{tip,tip}^{he})\mu_{tip}\nonumber \\ &
& +\frac{e^2}{h}(T_{tip,1}^{he}-T_{tip,1}^{ee})\mu_1\nonumber
\\ & = &
\frac{e^2}{h}(T_{1,tip}^{he}+T_{1,tip}^{ee}+2T_{tip,tip}^{he})
\mu_{tip}\nonumber \\ & &
+\frac{e^2}{h}(T_{tip,1}^{he}-T_{tip,1}^{ee})\mu_1\, ,
\label{stipcur.5}
\end{eqnarray}
where we used the unitarity of the scattering matrix in the second
step. In the absence of a magnetic field, we can use the
identities, Eqs.\ (\ref{symrel1.5})-(\ref{symrel2.5}) to replace
the emissivities by the injectivities. In addition, we see that
the Andreev reflection probability $T_{tip,tip}^{he}$ is of the
order $|t|^4$ so that we can neglect it in (\ref{stipcur.5}). With
the abbreviation $G=(e^2/h)4\pi^2\nu_{tip}|t|^2\nu(x)$ we can then
write
\begin{equation}
\langle I_{tip}\rangle
=G\left(\mu_{tip}-\frac{q(x)}{p(x)}\mu_1\right)\, .
\end{equation}
Here, we introduced the particle density $p(x)$ and the charge
density $q(x)$ by
\begin{eqnarray}
p(x) & = & \nu(x_e,1_e)+\nu(x_h,1_e)\, ,\\
 q(x) & = &
\nu(x_e,1_e)-\nu(x_h,1_e)\, .
\end{eqnarray}
Note, that in this definition the charge density does not contain
a factor $e$ but does only count the electron injectivity positive
and the hole injectivity negative. Using the tip as a perfect
voltage probe, i.~e.\ setting $I_{tip}=0$, gives
\begin{equation}
\mu_{tip}=\frac{q(x)}{p(x)}\mu_1\, .\label{messpot.5}
\end{equation}
If the hole-like injectivity is zero, i.~e.\ the electrons don't
see the superconductor, the tunneling tip measures the
electro-chemical potential $\mu_1$ of contact 1. In the following
chapter we will discuss further examples.
\subsection{Examples}
We use Eq.\ (\ref{messpot.5}) to investigate some $N-S$ structures
where one normal contact is connected to a superconductor. The
interface between the normal conductor and the superconductor is
always perfectly transparent, while the normal conductor can be
ballistic or contain scattering regions.

First we consider a perfect ballistic normal conductor. Since all
electrons that enter the conductor propagate to the
superconducting interface, where they are Andreev reflected as
holes, the electron- and hole-injectivities are everywhere in the
ballistic region identical. We have $\nu(x_e,1_e)=\nu(x_h,1_e)$
and therefore the measured potential at the tip $\mu_{tip}$ is
always zero, i.\ e.\ equal to the chemical potential of the
superconductor.

Next, we consider a normal one-channel conductor which contains a
scattering region. The scattering region is characterized by a
scattering matrix leading to the transmission probability $T$ and
reflection probability $R$. In between the scattering region and
the interface to the superconductor electron- and hole-injectivity
are the same, so that a voltage probe measures in this region
always the potential of the superconductor. To the left of the
scattering region, away from the superconductor, the electron- and
hole-injectivities are given by the normal- and Andreev-reflection
probabilities,
\begin{equation}
\nu(x_e,1_e)=1+T^{ee}\quad{\rm
and}\quad\nu(x_h,1_e)=T^{he},\label{scattinj.5}
\end{equation}
where the Andreev-reflection probability is
\begin{equation}
T^{he}=\frac{T^2}{(2-T)^2}
\end{equation}
and the normal reflection is $T^{ee}=1-T^{he}$. In calculating the
injectivities above we neglected the fast oscillating interference
terms between incoming and reflected waves. Using these
injectivities, the measured potential, Eq.\ (\ref{messpot.5}), to
the left of the scatterer is
\begin{equation}
\mu_{tip}=T^{ee}\mu_1=\frac{4(1-T)}{(2-T)^2}\mu_1\, .
\end{equation}
The voltage drop from the normal reservoir to the left side of the
scattering region is $\mu_1-\mu_{tip}=T^{he}\mu_1$. If we divide
this voltage drop by the current $I=2(e/h)T^{he}\mu_1$ flowing
through the sample we get the contact resistance
\begin{equation}
{\cal R}_C=\frac{\mu_1-\mu_{tip}}{eI}=\frac{h}{2e^2}\, .
\end{equation}
Since one measures on the right side of the scattering region
always the chemical potential of the superconductor, there is no
contact resistance on the superconducting side. The total contact
resistance of an $N-S$ structure is therefore half as big as the
one for a purely normal system. This is in agreement with the
findings in Refs.\ \cite{datta.5} and \cite{sols.5}.

As a last example we consider a metallic diffusive conductor in
the ensemble average. The diffusive region extends from $x=0$ to
$x=L$, where it is in contact with a superconductor. Its length
$L$ is much bigger than its width $W$ so that it is justified to
treat the diffusion to be effectively one-dimensional in the
ensemble average. We have to find the electron- and hole
injectivity. For this we have to solve the differential equations
\begin{equation}
\frac{d^2}{dx^2}\nu(x_e,1_e)=0\quad{\rm
and}\quad\frac{d^2}{dx^2}\nu(x_h,1_e)=0
\end{equation}
with the boundary conditions $\nu(x=0_e,1_e)=\nu_0$ and
$\nu(x=0_h,1_e)=0$, where $\nu_0=m^\star/2\pi\hbar^2$ is the two
dimensional density of states. In addition, due to the transparent
interface and the perfect Andreev reflection we have
$\nu(x=L_e,1_e)=\nu(x=L_h,1_e)$ and due to particle number
conservation we have $\nu(x_e,1_e)+\nu(x_h,1_e)=\nu_0$. Solving
the differential equations using the boundary conditions we get
the injectivities
\begin{equation}
\nu(x_e,1_e)=\nu_0\left(1-\frac{x}{2L}\right)\quad{\rm
and}\quad\nu(x_h,1_e)=\nu_0\frac{x}{2L}\, .\label{diffinj.5}
\end{equation}
Note here, that these densities look like the densities of a
diffusive conductor of length $2L$ connected to two normal
contacts \cite{gramespacher99}. The potential $\mu_{tip}(x)$
measured at a position $x$ along the diffusive conductor is
\begin{equation}
\mu_{tip}(x)=\left(1-\frac{x}{L}\right)\mu_1\, ,
\end{equation}
which means, that there is a linear voltage drop along the
diffusive conductor from $\mu_1$ on the left side to zero, the
electro-chemical potential of the superconductor, at the
interface. Thus, attaching a superconductor to a normal diffusive
wire does not change the voltage measurement along the wire.
\section{Current fluctuations and positive correlations}
In this section we investigate the time dependent current measured
at a normal contact in the presence of superconductivity. The
system we consider consists of several normal contacts and one
superconducting contact. In order to describe the time dependent
current it is useful to use the formulation of second
quantization. We closely follow the lines of Ref.\
\cite{buttiker92.5}, where the theory was developed for purely
normal conducting systems. The essential point in the presence of
superconductivity is, that at energies which are small compared to
the superconducting gap, there are no propagating quasiparticles
inside the superconductor. The superconductor influences the
normal part of the system only by reflecting electrons as holes
and vice versa and generating a supercurrent inside the
superconductor. A similar derivation of current correlations for
multiterminal $N-S$ hybrid structures has already been given by
Anantram and Datta in Ref.\ \cite{anantram.5}. Statistical
particle counting arguments have been used by Martin in
\cite{martin.5}.

First we need the current operator $\hat{I}_\alpha(t)$ for the
current in lead $\alpha$. The current consists of two parts, the
current carried by the electrons, $\hat{I}_\alpha^e(t)$, and the
current carried by the holes, $\hat{I}_\alpha^h(t)$. The total
current is the sum of these two currents which are given by
\begin{eqnarray}
\hat{I}_\alpha^\mu(t)=\frac{q^\mu}{h}\!\!\int_0^\infty\!\!\!\!\!\!
dEdE'\!\sum_{\beta\nu \atop \gamma\lambda}\! & & (\hat{\bf
a}_\beta^\nu)^\dagger (E) {\bf
A}_{\beta\nu,\gamma\lambda}(\alpha\mu;E,E')\hat{\bf
a}_\gamma^\lambda (E') \nonumber \\ & & \times \exp [
i(E-E')t/\hbar]
\end{eqnarray}
with the current matrix
\begin{equation}
{\bf
A}_{\beta\nu,\gamma\lambda}(\alpha\mu;E,E')=
\delta_{\beta\nu,\alpha\mu}\delta_{\gamma\lambda,\alpha\mu}
-({\bf s}_{\alpha\beta}^{\mu\nu})^\dagger(E){\bf
s}_{\alpha\gamma}^{\mu\lambda}(E')\, .
\end{equation}
The indices $\alpha,\beta$ and $\gamma$ enumerate the normal leads
and the indices $\mu,\nu,\lambda=e,h$. All energies have to be
measured relative to the electro-chemical potential of the
superconductor. We can now continue like in Ref.\
\cite{buttiker92.5} to find the low-frequency spectrum of the
current correlations\cite{anantram.5},
\begin{eqnarray}
\langle \Delta I_\alpha^\mu\Delta I_\beta^\nu\rangle=2\frac{q^\mu
q^\nu}{h} \sum_{\gamma\lambda\atop\delta\kappa}\!\int_0^\infty
\!\!\!\!\!\! dE & &{\rm Tr}\left[ {\bf
A}_{\gamma\lambda,\delta\kappa}(\alpha\mu){\bf
A}_{\delta\kappa,\gamma\lambda}(\beta\nu)\right]\nonumber \\ & &
\times f_\gamma^\lambda(E)( 1-f_\delta^\kappa(E))\, .
\label{gencorfluc.5}
\end{eqnarray}
In this expression both energy arguments in the current matrix are
equal to $E$ and the distribution functions are the ones given in
Eq.\ (\ref{distfunct}). Eq.\ (\ref{gencorfluc.5}) is in fact
equivalent to Eq.\ (1.16) of Ref.\ \cite{buttiker92.5} if we keep
in mind that, in the superconducting case, we replaced the
scattering matrix between $N+1$ normal contacts by an effective
scattering matrix between $2N$ contacts (contacts for electrons
and holes counted separately) labelled by pairs of indices
$\alpha\mu$. As the normal $(N+1)\times (N+1)$ scattering matrix,
the effective $2N\times 2N$ scattering matrix is unitary. The
integral in Eq.\ (\ref{gencorfluc.5}) goes only from zero to
infinity because we add particle and hole excitations.

Now we can apply this formula to an $N-S$ system which has an
additional local tunneling contact, called tip (c.\ f.\ figure
\ref{nsfig.5}). The fluctuations of the currents at the tip are
\begin{equation}
\langle (\Delta I_{tip})^2\rangle=\langle (\Delta
I_{tip}^e)^2\rangle +\langle (\Delta I_{tip}^h)^2\rangle+2\langle
\Delta I_{tip}^e \Delta I_{tip}^h \rangle\, .
\end{equation}
We have to evaluate the current matrix using the effective
scattering matrix for this system. To the lowest order in the
coupling strength $t$ of the tip this gives
\begin{eqnarray}
\langle (\Delta I_{tip})^2\rangle & = &
2\frac{e^2}{h}4\pi^2\nu_{tip}|t|^2\left[
\nu(x_e,1_e)(\mu_1-\mu_{tip})\right. \nonumber \\ & + & \left.
\nu(x_e,1_h)\mu_{tip}+\nu(x_h,1_e) \mu_1\right] \, .
\end{eqnarray}
If we chose the electrochemical potential at the tip according to
Eq.\ (\ref{messpot.5}) such that there is on average zero current
flowing into the tip, we can write for the current fluctuations at
the tip
\begin{equation}
\langle (\Delta I_{tip})^2\rangle=8eG_0V
\frac{\nu(x_e,1_e)}{\nu(x)}\left(1-\frac{\nu(x_e,1_e)}{\nu(x)}
\right) \label{deltaisn.5}
\end{equation}
with $G_0=(e^2/h)4\pi^2\nu_{tip}|t|^2\nu(x)$ and the applied bias
$eV=\mu_1$.
\subsection{Examples of current fluctuations}
In a ballistic conductor electron- and hole-injectivity are
identical so that the fluctuations at the tip become,
\begin{equation}
\langle (\Delta I_{tip})^2\rangle=2eG_0V\, .
\end{equation}
We compare this result with the case where the superconductor is
replaced by a normal conductor found in \cite{gramespacher99},
$\langle(\Delta I_{tip})^2\rangle=eG_0^NV$, with the conductance
$G_0^N=(e^2/h)4\pi^2\nu_{tip}|t|^2\left[\nu(x,1)+\nu(x,0)\right]$.
Switching on superconductivity thus doubled the noise measured at
the weakly coupled contact. A doubling of the fluctuations of the
current at a normal ballistic wire which is connected to a
superconductor is also found in \cite{dejong.5} and
\cite{muzykanskii}.

The next example is a normal conductor with a scatterer of
transmission probability $T$. Using the electron- and
hole-injectivity for such a system to the left of the scatterer
and neglecting phase coherence, Eqs.\ (\ref{scattinj.5}), gives
the noise spectrum,
\begin{equation}
\langle (\Delta I_{tip})^2\rangle=2eG_0VT^{he}(2-T^{he})\, .
\end{equation}
Again, comparing this expression to the result for a normal wire
\cite{gramespacher99} shows the appearance of a factor of two in
the superconducting case and the appearance of the Andreev
reflection probability $T^{he}$ instead of the normal transmission
probability $T$.

The last and most interesting example is the metallic diffusive
wire. Inserting the corresponding injectivities, Eqs.\
(\ref{diffinj.5}), into the expression for the current
fluctuations yields
\begin{equation}
\langle (\Delta
I_{tip})^2\rangle=8eG_0V\frac{x}{2L}\left(1-\frac{x}{2L}\right)\,
.
\end{equation}
This can be compared with the one obtained for a purely normal
conductor, $\langle(\Delta
I_{tip})^2\rangle=4eGV\frac{x}{L}(1-\frac{x}{L})$. Whereas the
voltage measurement showed the same linear voltage drop over a
diffusive region in the purely normal case as well as in the case
where the diffusive part is in contact with a superconductor, the
fluctuation spectrum measured at the tip shows a different
behaviour. The fluctuations are maximal at the interface $x=L$ and
vanish at the contact to the normal reservoir $x=0$. This can be
understood, if we think of the interface to the superconductor as
a mirror thus representing the middle of a fictitious wire of
length $2L$ which continues after the interface into the
superconductor \cite{brouwer.5}. However, at the interface to the
superconductor, the measured fluctuations are $\langle (\Delta
I_{tip})^2\rangle=2eG_0V$, twice as large as the fluctuations
measured in the middle of a purely normal wire. We could also
compare the normal wire of length $2L$ with a wire of length $L$
connected to a superconductor. Then, on the first half of the
purely normal wire ($0<x<L$), the correlations of the purely
normal and the hybrid structure differ in exactly a factor of two.
\subsection{Examples of current cross-correlations}
In this section we investigate the correlations of the currents
measured at two normal contacts of a normal-superconducting hybrid
structure starting from equation (\ref{gencorfluc.5}). The system
we consider is again shown in figure \ref{nsfig.5}. For purely
normal systems, the correlations of fermions are due to their
exclusion statistic always negative. However, the presence of a
nearby superconductor induces bosonic correlations between the
electrons in a normal conductor. It was therefore shown by
analytic calculations in \cite{anantram.5,martin.5} that the
bosonic character of electron pairs can in principle lead to
positive correlations of the currents at two normal contacts. In
\cite{anantram.5} one could in fact find through numerical
investigation positive correlations at two normal contacts, which
were sandwiched in between a ring shaped superconductor
representing the third terminal. Very recently, positive
correlations were found for a system consisting of a wave splitter
connected to a superconductor\cite{torres.5}. We show here in an
analytic calculation that one can get positive correlations on a
system where one of the two normal contacts is only weakly
coupled. The calculation presented below sheds light on the truely
mesoscopic origin of these positive correlations. The existence of
positive correlations is in contrast to the recently measured
negative correlations on purely normal conductors by Henny {\em et
al}.\cite{henny} and Oliver {\em et al}.\cite{oliver}.

The contact 1 is held at the same potential as the superconductor
whereas a bias $V$ is applied at the tip. We restrict ourselves to
the case of zero temperature. Then only the Fermi distribution
function $f^e_2(E)$ of the electrons in the tip is different from
zero,
\begin{equation}
f_2^e(E)=\theta(eV -E)\, ,
\end{equation}
all other distribution functions vanish (One has to keep in mind
that we are only interested in the distribution function for
energies which are larger than the electro-chemical potential of
the superconductor). We have now to evaluate the general formula
for the current-correlations using these Fermi functions. We
expand the effective scattering matrix for scattering between the
two normal contacts in powers of the coupling energy $t$ and get
the following correlations,
\begin{eqnarray}
\langle\Delta I_1^e\Delta I_{tip}^e\rangle & = & -\alpha
4\pi^2\nu_{tip}|t|^2\nu(x_e,1_e)\, , \\ \langle\Delta I_1^h\Delta
I_{tip}^e\rangle & = & \alpha 4\pi^2\nu_{tip}|t|^2\nu(x_e,1_h)\, ,
\\ \langle\Delta I_1^e\Delta I_{tip}^h\rangle & = & \langle\Delta
I_1^h\Delta I_{tip}^h\rangle = 0\, ,
\end{eqnarray}
with $\alpha=2\frac{e^2}{h}eV$. The total correlation of the
currents at contact 1 and 2 is the sum of all four terms. In the
absence of a magnetic field, the correlations are proportional to
the injected charge density $q(x)=\nu(x_e,1_e)-\nu(x_h,1_e)$,
\begin{eqnarray}
\langle\Delta I_1\Delta I_{tip}\rangle & = & -2e\frac{e^2}{h}V
4\pi^2\nu_{tip}|t|^2 q(x) \nonumber \\ & = &
-2eG_0V\frac{q(x)}{p(x)}
\end{eqnarray}
with the conductance $G_0$ defined as for equation
(\ref{deltaisn.5}). This means, that if at the point $x$ the hole
density of in contact 1 injected electrons is larger than the
electron density, the net injected charge density becomes negative
and therefore, the correlations become positive. The electron and
hole injectivity are proportional to the absolute squared value of
the corresponding wavefunction amplitude. For a system consisting
of a barrier in the normal conducting part the wavefunction of an
injected electron in contact 1 is to the left of the barrier given
by
\begin{eqnarray}
\psi_{1e}(x_e) & = & e^{ikx}+r^{ee}e^{-ikx}\, , \\ \psi_{1e}(x_h)
& = & r^{he}e^{-ikx}\, .
\end{eqnarray}
Here, $r^{ee}=\sqrt{R^{ee}}$ is the normal reflection amplitude
and $r^{he}=\sqrt{R^{he}}$ is the Andreev reflection amplitude.
For simplicity, we chose these two amplitudes to be real. The
injected charge density at a point $x$ is then
\begin{equation}
q(x)=\frac{2}{hv}\left\{R^{ee}+\sqrt{R^{ee}}\cos(2kx)\right\}\, .
\end{equation}
One sees immediately, that, since $0\le R^{ee}\le 1$, one can find
for every $R^{ee}$ positions where the charge density is negative
and the measured correlations become positive. Note, however, that
this is only the case if one respects the phase coherence of
incoming and reflected wave (as it is done in the above
calculation). If phase coherence is neglected, the cosine term
disappears and the injected charge is always positive. (Remember
that we defined the charge density $q(x)$ such that it is positive
if there is a net electronic charge in the system and negative if
there is a net hole charge.) Similarly, the cosine term averages
out if one considers a phase-coherent conductor with $N\gg 1$ open
channels. If we assume to have no interchannel scattering and that
the reflection amplitudes $R_m^{ee}\equiv R$ are independent of
the channel index $m$, the charge density at a point $x$ is
\begin{equation}
q(x)=\sum_{m=1}^N\frac{2}{hv_m}\left[R+\sqrt{R}\cos(2k_m
x-\delta_m)\right]\, ,
\end{equation}
with the scattering phase shifts $\delta_m$. It is easily seen,
that the cosine terms will cancel out for randomly distributed
phase shifts, thus destroying the possibility of positive
correlations. Neither does one get positive correlations for a
metallic diffusive wire in the ensemble average using the above
derived densities. Positive correlations are therefore, at least
in the setup described here, a truly mesoscopic effect and can
only be expected to be seen on one or few channel conductors that
preserve the phase coherence. Up to now all examples exhibiting
positive correlations \cite{anantram.5,martin.5,torres.5} were
perfectly phase coherent one channel conductors in the
neighborhood of a superconductor. Furthermore, it is instructive
to note that Ref.\ \cite{anantram.5} finds a sign change of the
correlations as a function of the Aharonov-Bohm flux, which
determines the phase difference at the two interfaces to the
superconductor. This also points to a purely mesoscopic effect.
Thus the picture which emerges is that in hybrid systems there is
a mesoscopic effect of order $1/N$, which can give a positive
contribution in correlations. The main part (of order 1) has the
sign observed in normal conductors.
\section{Conclusions}
We have used the effective scattering matrix to define fundamental
partial densities of states, Eq.\ (\ref{fondpdos}), for systems
containing normal and superconducting parts. These partial
densities and the injectivities and emissivities constructed from
them are shown to be very useful in the description of current and
current fluctuations and correlations measured at the normal
contacts of a hybrid structure. The current at a tunnteling tip is
sensitive to the effective distributions of electrons and holes
inside a normal multiprobe conductor to which one superconducting
reservoir is attached. The local distributions are evaluated for
conductors containing a single scatterer and for disordered
conductors in the ensemble average. We also gave an example
illustrating the possibility of positive correlations of the
currents at two normal contacts due to induced bosonic behaviour.
However, we reasoned that positive correlations are a mesoscopic
($1/N$) effect, which experimentally can only be expected to be
observable on conductors containing very few open channels.
\section{Acknowledgment}
We thank Andrew M.\ Martin for useful discussions. This work was
supported by the Swiss National Science foundation.
\vspace{-0.5cm}

\end{multicols}

\begin{references}
%
\vspace{-1.2cm}
%
\bibitem{andreev}
A.\ F.\ Andreev, Zh.\ Eksp.\ teor.\ Fiz.\ {\bf 46}, 1823 (1964)
[Sov.\ Phys.\ JETP {\bf 19}, 1228 (1964)].
%
\bibitem{blonder.5}
G.\ E.\ Blonder, M.\ Tinkham, and T.\ M.\ Klapwijk, Phys.\ Rev.\ B
{\bf 25}, 4515 (1982).
%
\bibitem{petrashov}
V.\ T.\ Petrashov,V.\ N.\ Antonov, P.\ Delsing, and T.\
Claeson, Phys.\ Rev.\ Lett.\ {\bf 74}, 5268 (1995).
%
\bibitem{takayanagi}
H.\ Takayanagi, T.\ Akazaki, and J.\ Nitta, Phys.\ Rev.\ Lett.\
{\bf 75}, 3533 (1995).
%
\bibitem{courtois}
H.\ Courtois, Ph.\ Gandit, D.\ Mailly, and B.\ Pannetier,
Phys.\ Rev.\ Lett.\ {\bf 76}, 130 (1996).
%
\bibitem{gueron}
S.\ Gu\'eron, H.\ Pothier, N.\ O.\ Birge, D.\ Esteve, and M.\ H.\
Devoret, Phys.\ Rev.\ Lett.\ {\bf 77}, 3025 (1996).
%
\bibitem{hartog}
S.\ G.\ den Hartog, C.\ M.\ A.\ Kapteyn, B.\ J.\ van Wees, T.\ M.\
Klapwijk, and G.\ Borghs, Phys.\ Rev.\ Lett.\ {\bf 77}, 4954
(1996).
%
\bibitem{beenakkerreview}
For reviews see, e.\ g., C.\ W.\ J.\ Beenakker, Rev.\ Mod.\ Phys.\
{\bf 69}, 731 (1997); C.\ J.\ Lambert and R.\ Raimondi, J.\ Phys.:
Condens.\ Matter {\bf 10}, 901 (1998) and references therein.
%
%\bibitem{belzig}
%W.\ Belzig, C.\ Bruder, and G.\ Sch\"on, Phys.\ Rev.\ B {\bf 54},
%9443 (1996).
%
%\bibitem{zhou}
%F.\ Zhou and B.\ Spivak, Phys.\ Rev.\ Lett.\ {\bf 80}, 3847
%(1998).
%
\bibitem{wilhelm}
W.\ Belzig, F.\ K.\ Wilhelm, C.\ Bruder, G.\ Sch\"on, and A.\ D.\
Zaikin, Superlattices and Microstructures {\bf 25}, 1251 (1999).
%
\bibitem{gramespacherthesis}
T.\ Gramespacher, Ph.D.\ Thesis, University of Geneva,
Switzerland, 1999.
%
\bibitem{anantram.5}
M.\ P.\ Anantram and S.\ Datta, Phys.\ Rev.\ B {\bf 53}, 16 390
(1996).
%
\bibitem{martin.5}
T.\ Martin, Physics Letters A {\bf 220}, 137 (1996).
%
\bibitem{torres.5}
J.\ Torr\`es and T.\ Martin, cond-mat/9906012.
%
\bibitem{mortensen}
N.\ A.\ Mortensen, A.-P.\ Jauho, and K.\ Flensberg, (unpublished).
%
\bibitem{buttiker85}
M.\ B\"uttiker, Phys.\ Rev.\ B {\bf 32}, 1846 (1985).
%
\bibitem{weidenmueller}
S.\ Iida and H.\ A.\ Weidenm\"uller, Phys.\ Rev.\ Lett.\ {\bf 64},
583 (1990); Ann.\ Phys.\ (N.Y.) {\bf 200} 219 (1990).
%
\bibitem{frahm.5}
K.\ M.\ Frahm, P.\ W.\ Brouwer, J.\ A.\ Melsen, and C.\ W.\ J.\
Beenakker, Phys.\ Rev.\ Lett.\ {\bf 76}, 2981 (1996).
%
\bibitem{gasparian}
V.\ Gasparian, T.\ Christen, and M.\ B\"uttiker, Phys.\ Rev.\ A
{\bf 54}, 4022 (1996).
%
\bibitem{martin}
A.\ M.\ Martin, T.\ Gramespacher, and M.\ B\"uttiker, Phys.\ Rev.\
B, (1999); cond-mat/9907240.
%
\bibitem{lambert}
C.\ J.\ Lambert, V.\ C.\ Hui, and S.\ J.\ Robinson, J.\ Phys.:
Condens.\ Matter {\bf 5}, 4187 (1993).
%
\bibitem{buttiker92.5}
M.\ B\"uttiker, Phys.\ Rev.\ B {\bf 46}, 12 485 (1992).
%
\bibitem{gramespacher99}
T.\ Gramespacher and M.\ B\"uttiker, Phys.\ Rev.\ B {\bf 60}, 2375
(1999).
%
\bibitem{datta.5}
S.\ Datta, P.\ F.\ Bagwell, and M.\ P.\ Anantram,
Phys.\ Low-Dim.\ Struct.\ {\bf 3}, 1 (1996).
%
\bibitem{sols.5}
F.\ Sols and J.\ S\'anchez-Ca\~nizares, Superlattices and
Microstructures {\bf 25}, 627 (1999).
%
\bibitem{dejong.5}
M.\ J.\ M.\ de Jong and C.\ W.\ J.\ Beenakker, Phys.\ Rev.\ B {\bf
49}, 16 070 (1994).
%
\bibitem{muzykanskii}
B.\ A.\ Muzykanskii and D.\ E.\ Khmelnitskii, Phys.\ Rev.\ B {\bf
50}, 3982 (1994).
%
\bibitem{brouwer.5}
P.\ W.\ Brouwer and C.\ W.\ J.\ Beenakker, Phys.\ Rev.\ B {\bf
52}, 16 772 (1995).
%
\bibitem{henny}
M.\ Henny, S.\ Oberholzer, C.\ Strunk, T. Heinzel, K. Ensslin, M.
Holland, and C. Schönenberger, Science {\bf 284}, 296 (1999).
%
\bibitem{oliver}
W.\ D.\ Oliver, J.\ Kim, R.\ C.\ Liu, and Y.\ Yamamoto, Science
{\bf 284}, 299 (1999).
%
\end{references}
\end{document}